# DETECTION OF LOW ENERGY ELECTRONS BY A TOTAL ABSORPTION CHERENKOV SPECTROMETER

V.I. Alekseev[a], V.A. Baskov[a]*, V.A. Dronov[a], A.I. L'vov[a], A.V. Koltsov[a], Yu.F. Krechetov[b], E.I. Malinovsky[a], V.V. Polyansky[a]

*[a] P.N. Lebedev Physical Institute, Moscow, 119991 Russia*

*[b] Joint Institute for Nuclear Research, Dubna, Moscow Region, 141980 Russia*

*\*E-mail: baskov@x4u.lebedev.ru*

Presented are results of calibration, with a quasi-monochromatic electron beam, of a total absorption Cherenkov spectrometer based on the 14.8 $X_0$ lead glass TF-1. It was found that the energy resolution of the spectrometer was 89% to 10% in the energy range of the electron beam $E$ = 6 to 285 MeV, respectively.

The electron synchrotron S-25R ("Pakhra") [1] of the Lebedev Physical Institute of the Russian Academy of Sciences (LPI) is an accelerator currently generating external beams of photons, electrons and positrons with energies up to 850 MeV. A calibration channel of secondary quasi-monochromatic electrons with energies between 6 and 300 MeV and intensity of ~100 electron/s was created at the bremsstrahlung photon beam of this accelerator [2, 3].

Energy characteristics of the calibration electron beam were determined, in addition to other methods, by a total absorption Cherenkov spectrometer (TAChS) based on the lead glass TF-1 with the thickness of 14.8 radiation lengths $X_0$ ($X_0$ = 2.5 cm, $\rho$ = 3.86 g/cm$^3$) [3-5].

The choice of TAChS as a detector was determined by the fact that such Cherenkov spectrometers make it possible to detect photons and electrons/positrons and to determine their energies in a wide energy range from several MeV to tens of GeV [5] due to:

1. Threshold of radiation of electrons in a dense matter is about hundreds of keV;

2. Sufficiently high energy resolution (about 25 to 5%) in a wide energy range of 0.1 to 100 GeV, respectively;

3. A linear relation between the detected particle energy and the output amplitude of the spectrometer;

4. With the growth of the energy of electrons and photons, the thickness of the radiator, providing full energy absorption, increases proportionally to the logarithm of the energy and is about 15 $X_0$ at energies up to 0.5 GeV;

5. Cherenkov radiation features allow to receive nanosecond pulses at the output of photodetectors that makes it possible to use signals from the spectrometer in fast electronic circuits.

In order to determine energy characteristics of the calibration beam, a TAChS is used that was very successful in previous works with the secondary electron beam of the proton accelerator of the Institute for High Energy Physics (IHEP) in Protvino [4,5].

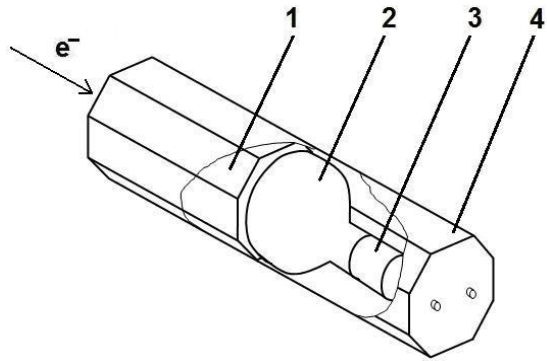

**Fig. 1.** Scheme of the total absorption Cherenkov spectrometer (TAChS):
1 = radiator, 2 = PMT-49, 3 = voltage divider, 4 = housing.

The built TAChS is a hexagonal radiator of the lead glass TF-1 with the inscribed circle diameter 18 cm and length 37 cm (Fig. 1). On all sides, except for the area occupied by a PMT photocathode, the radiator is wrapped in an aluminized lavsan. The TAChS uses PMT-49 with a standard divider [3-5].

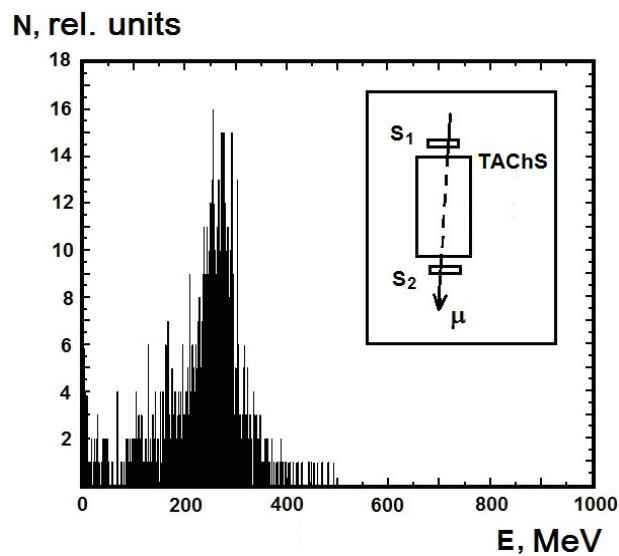

**Fig. 2.** Calibration spectrum of cosmic muons measured by the TAChS.
In the insert: $S_1$ and $S_2$ are scintillation counters.

The TAChS preliminary calibration was performed with cosmic-ray muons according to the method of "all through" presented in the inset of Fig. 2 [3]. The trigger was the signal from the coincidence of signals of two scintillation counters $S_1$ and $S_2$ having scintillators of $70 \times 70 \times 5$ mm$^3$. The average energy release of muons in the TAChS radiator was ≈285 MeV that corresponded to the 275th channel in the amplitude spectrum shown in Fig. 2 when the voltage across the power divider of PMT-49 was $U = 1600$ V. The coefficient of proportionality $k$ relating the average energy $E$ left by the cosmic muon in TAChS and the average amplitude $A$ in the spectrum ($E = k \cdot A$), except the pedestal, was $k \approx 1.04$ MeV/channel.

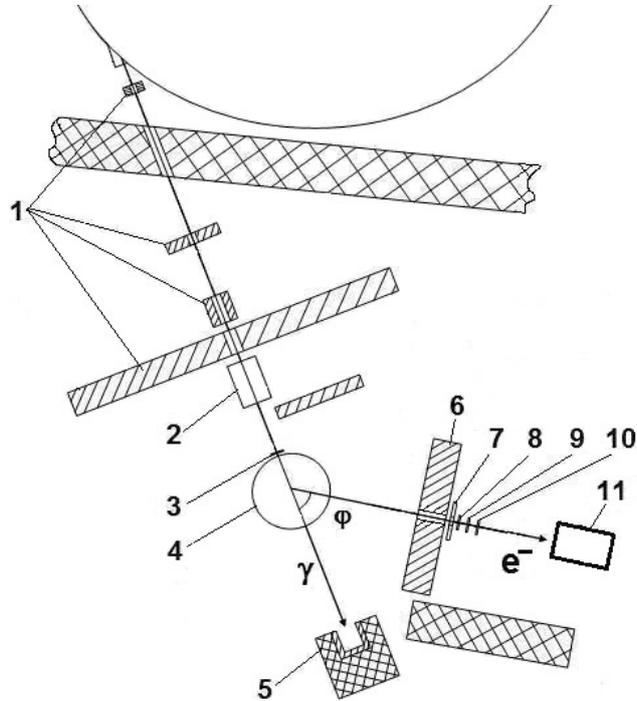

**Fig. 3.** The scheme of calibration of TAChS at a quasi-monochromatic electron beam:
1 = lead collimators; 2 = cleaning magnet SP-3; 3 = converter; 4 = magnet SP-57; 5 = photon beam absorber ("mortuary"); 6 = collimator (∅10 mm); 7 = scintillation veto-counter A; 8,9,10 = scintillation counters $S_1$, $S_2$, $S_3$; 11 = TAChS.

The basic calibration of the TAChS took place at the beam of quasi-monochromatic electrons of the accelerator «Pakhra» simultaneously with the study of its energy characteristics [3]. The electron energy of the primary beam in the accelerator chamber was 500 MeV. The calibration scheme is shown in Fig. 3. The electrons resulting from interaction of the bremsstrahlung photon beam with the converter (3) were deflected by the magnetic field of the magnet SP-57 (4) from the main trajectory of the photon beam. Electrons moving at the angle of φ = 36° got into a lead collimator (6) with an orifice of ∅10 mm and a thickness of 70 mm located at a distance of 3 m from the poles of the magnet SP-57. The trigger was formed by coincidence of the signals from the telescope of scintillation counters (7–10), T = $(S_1 \cdot S_2 \cdot S_3) \cdot A$ (A = veto-counter of 60×90×10 mm³, $S_1 - S_3$ = flip-flop counters of 15×15×1 mm³).

Figure 4 shows the dependence of TAChS average signal amplitudes on electron energies determined in units of channels of a charge-to-digital converter by constructing topographies of magnetic fields corresponding to set points of the current of the SP-57 magnet windings with the following calculation of the passage of electrons from the converter to the collimator. It is seen that in the entire range of the investigated electron energies the dependence is linear.

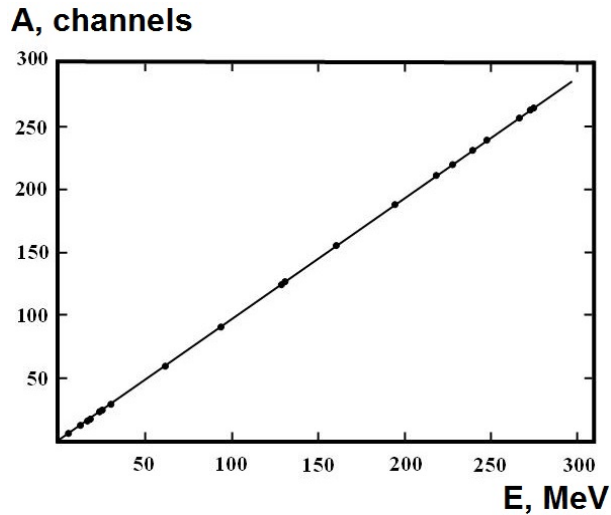

**Fig. 4.** Dependence of the amplitude A of the TAChS signal on the energy $E$ of the quasi-monochromatic electron beam.

The dependence of the TAChS energy resolution on the electron energy except the energy resolution of the electron beam [5] is shown in Fig. 5. It was found that the resolution of TAChS in the electron energy range of 30 – 300 MeV is described as

$$\Delta E/E = 0.015 + 0.05/\sqrt{E}, \qquad (1)$$

where $\Delta E$ and $E$ (in MeV) are the full width at half maximum of the energy spectrum and the average electron energy, respectively.

In the low-energy region of electrons of 6–30 MeV, the energy resolution of the TAChS is well described by the dependence

$$\Delta E/E = 0.015 + 0.068/\sqrt{E}. \qquad (2)$$

Deterioration in the TAChS resolution in this energy region is probably associated with an increase in fluctuations of photoelectrons on the PMT photocathode that in turn is associated with insufficient amount of Cherenkov light due to a decrease in the number of shower tracks at low energies of electrons.

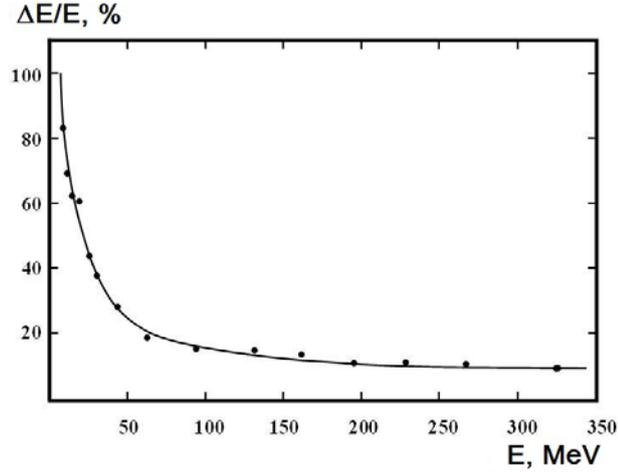

**Fig. 5.** Dependence of the energy resolution Δ*E*/*E* of TAChS
(Δ*E* and *E* being the full width at half maximum of the energy spectrum
and the average electron energy, respectively)
on the energy of the quasi-monochromatic electron beam (E).

Calibration of this TAChS at electron beams of energies of tens of GeV determined its energy resolution (without taking into account the energy resolution of the electron beam which, at an average electron pulse of 31 GeV/*c*, was equal to √DE =0.72 GeV/*c*, where DE is the dispersion of electron energy in the primary beam) [4,5]

$$\Delta E/E = 0.015 + 0.125/\sqrt{E}. \qquad (3)$$

At electron energies of 30 – 300 MeV, the formula (3), even taking into account the energy resolution of the electron beam, gives the values of the spectrometer resolution exceeding the experimental ones by ~2.5 times. This is probably because, in the volume of the spectrometer with 100% efficiency of electron detection, the energy of electrons remains completely inside the detector. This can be clearly seen from the pre-calibration of the TAChS, in which the total energy release of muons in TAChS is ≈ 285 MeV [3]. When increasing the energy of electrons to several GeV and above, due to increase in the geometric characteristics of electromagnetic showers, they will not fit into the volume of the spectrometer and the energy of showers will go partly through sides and back of the spectrometer. The effect of the shower partly coming out of the spectrometer should affect the relative energy resolution of the spectrometer.

Thus, the total absorption Cherenkov spectrometer has a high energy resolution not only at electron energies of tens of GeV but also in the energy range of ~300 MeV that makes it possible to use the spectrometer in studies of the energy characteristics of calibration electron beams. Satisfactory time characteristics of the Cherenkov radiation, with a planned replacement of the "slow" PMT-49 with "high-speed" ones like PMT-63 or XP2040, will make it possible to use the TAChS in studies of the energy characteristics of the intense ($10^4 – 10^6$ electron/s) electron beam from the accelerator «Pakhra».

This work was supported by grants from the Russian Foundation for Basic Research (NICA - RFBR) No. 18-02-40061 and No. 18-02-40079.